\documentclass[a4paper]{article}

\usepackage{icrc2013}
\usepackage[english]{babel}
\usepackage{amssymb}

\title{Expected performance of the ASTRI-SST-2M telescope prototype}

\shorttitle{ASTRI telescope prototype}

\authors{
C.~Bigongiari$^{1a}$,
F.~Di~Pierro$^{1a}$,
C.~Morello$^{1a}$,
A.~Stamerra$^{1a}$,
P.~Vallania$^{1a}$,
G.~Agnetta$^{2}$,
L.A.~Antonelli$^{3}$,
D.~Bastieri$^{4}$,
G.~Bellassai$^{5}$,
M.~Belluso$^{5}$,
S.~Billotta$^{5}$,
B.~Biondo$^{2}$,
G.~Bonanno$^{5}$,
G.~Bonnoli$^{6}$,
P.~Bruno$^{5}$,
A.~Bulgarelli$^{7}$,
R.~Canestrari$^{6}$,
M.~Capalbi$^{2}$,
P.~Caraveo$^{8}$,
A.~Carosi$^{3}$,
E.~Cascone$^{9}$,
O.~Catalano$^{2}$,
M.~Cereda$^{6}$,
P.~Conconi$^{6}$,
V.~Conforti$^{7}$,
G.~Cusumano$^{2}$,
V.~De~Caprio$^{9}$,
A.~De~Luca$^{8}$,
A.~Di~Paola$^{3}$,
D.~Fantinel$^{10}$,
M.~Fiorini$^{8}$,
D.~Fugazza$^{6}$,
D.~Gardiol$^{1b}$,
M.~Ghigo$^{6}$,
F.~Gianotti$^{7}$,
S.~Giarrusso$^{2}$,
E.~Giro$^{10}$,
A.~Grillo$^{5}$,
D.~Impiombato$^{2}$,
S.~Incorvaia$^{8}$,
A.~La~Barbera$^{2}$,
N.~La~Palombara$^{8}$,
V.~La~Parola$^{2}$,
G.~La Rosa$^{2}$,
L.~Lessio$^{10}$,
G.~Leto$^{5}$,
S.~Lombardi$^{3}$,
F.~Lucarelli$^{3}$,
M.C.~Maccarone$^{2}$,
G.~Malaguti$^{7}$,
G.~Malaspina$^{6}$,
V.~Mangano$^{2}$,
D.~Marano$^{5}$,
E.~Martinetti$^{5}$,
R.~Millul$^{6}$,
T.~Mineo$^{2}$,
A.~Mist\`{o}$^{6}$,
G.~Morlino$^{11}$,
M.R.~Panzera$^{6}$,
G.~Pareschi$^{6}$,
G.~Rodeghiero$^{10}$,
P.~Romano$^{2}$,
F.~Russo$^{2}$,
B.~Sacco$^{2}$,
N.~Sartore$^{8}$,
J.~Schwarz$^{6}$,
A.~Segreto$^{2}$,
G.~Sironi$^{6}$,
G.~Sottile$^{2}$,
E.~Strazzeri$^{2}$,
L.~Stringhetti$^{8}$,
G.~Tagliaferri$^{6}$,
V.~Testa$^{3}$,
M.C.~Timpanaro$^{5}$,
G.~Toso$^{8}$,
G.~Tosti$^{12}$,
M.~Trifoglio$^{7}$,
S.~Vercellone$^{2}$,
V.~Zitelli$^{13}$
(the ASTRI Collaboration), 
L.~Arrabito$^{14}$,
K.~Bernl\"ohr$^{15,16}$,
G.~Maier$^{17}$,
N.~Komin$^{18}$
and the CTA Consortium.
}

\afiliations{
$^{1}$  INAF - Osservatorio Astrofisico di Torino; (a) Sede di Torino - Via P. Giuria 1, 10125 Torino, Italy; (b) Sede di Pino Torinese - Strada Osservatorio 20, 10025 Pino Torinese (TO), Italy\\
$^{2}$  INAF - Istituto di Astrofisica Spaziale e Fisica Cosmica di Palermo, Via U. La Malfa 153, 90146 Palermo, Italy\\
$^{3}$  INAF - Osservatorio Astronomico di Roma, Via Frascati 33, 00040 Monte Porziocatone (RM), Italy\\
$^{4}$  Universit\`{a} di Padova, Dip. Fisica e Astronomia, Via Marzolo 8, 35131 Padova, Italy\\
$^{5}$  INAF - Osservatorio Astrofisico di Catania, Via S. Sofia 78, 95123 Catania, Italy\\
$^{6}$  INAF - Osservatorio Astronomico di Brera, Via E. Bianchi 46, 23807 Merate (LC), Italy\\
$^{7}$  INAF - Istituto di Astrofisica Spaziale e Fisica Cosmica di Bologna, Via P. Gobetti 101, 40129 Bologna, Italy\\
$^{8}$  INAF - Istituto di Astrofisica Spaziale e Fisica Cosmica di Milano, Via E. Bassini 15, 20133 Milano, Italy\\
$^{9}$  INAF - Osservatorio Astronomico di Capodimonte, Salita Moiariello 16, 80131 Napoli, Italy\\
$^{10}$ INAF - Osservatorio Astronomico di Padova, Vicolo Osservatorio 5, 35122 Padova, Italy\\
$^{11}$ INAF - Osservatorio Astrofisico di Arcetri, Largo E. Fermi 5, 50125 Firenze, Italy\\
$^{12}$ Universit\`{a} di Perugia, Dip. Fisica, Via A. Pascoli, 06123 Perugia, Italy\\
$^{13}$ INAF - Osservatorio Astronomico di Bologna, Via Ranzani 1, 40127 Bologna, Italy\\
$^{14}$ Laboratoire Univers et Particules, Universit\'e de Montpellier II Place Eug\'ene Bataillon - CC 72, CNRS/IN2P3, F-34095 Montpellier, France \\
$^{15}$ Max-Planck-Institut f\"ur Kernphysik, P.O.Box 103980, D-69029 Heidelberg, Germany \\
$^{16}$ Institut f\"ur Physik, Humboldt-Universit\"at zu Berlin, Newtonstr. 15, D-12489 Berlin, Germany\\
$^{17}$ DESY, Platanenallee 6, D-15738 Zeuthen, Germany\\
$^{18}$ Laboratoire d'Annecy-le-Vieux de Physique des Particules, Universit\'e de Savoie, CNRS/IN2P3, France\\
}

\email{ciro.bigongiari@to.infn.it}

\abstract{

ASTRI (Astrofisica con Specchi a Tecnologia Replicante Italiana) 
is an Italian flagship project pursued by INAF (Istituto Nazionale di 
Astrofisica) strictly linked to the development of the Cherenkov Telescope Array, CTA. 
Primary goal of the ASTRI program is the design and production of an end-to-end prototype 
of a Small Size Telescope for the CTA sub-array devoted to the highest gamma-ray energy region. 
The prototype, named ASTRI SST-2M, will be tested on field in Italy during 2014.
This telescope will be the 
first 
Cherenkov telescope
adopting the double reflection layout in a 
Schwarzschild-Couder configuration with a tessellated primary mirror 
and a monolithic secondary mirror. 
The collected 
light will be focused on a compact and light-weight camera based on 
silicon photo-multipliers covering a $9.6^{\circ}$ 
full field of view. 
Detailed Monte Carlo simulations have been performed to estimate the performance 
of the planned telescope. The results regarding its energy threshold, 
sensitivity and angular resolution are shown and discussed.

}

\keywords{Gamma-ray astronomy, Cherenkov telescope, ASTRI, CTA.}

\begin{document}
\maketitle
%
%
\section{Introduction}
ASTRI (Astrofisica con Specchi a Tecnologia Replicante Italiana)
\cite{bib:ASTRI1,bib:ASTRI2} 
is an INAF flagship project dedicated to the development 
of the next generation of IACTs within the framework of the CTA 
(Cherenkov Telescope Array \cite{bib:CTA1,bib:CTA2}) International Observatory. 
In this context, INAF contribution is mainly focused on the high energy sub-array,
composed of up to 70 
Small Size Telescopes (SST) and dedicated  
to study the primary gamma-ray spectrum in the energy range between $\sim 1$~TeV and beyond 100~TeVs. 
For this purpose the ASTRI Collaboration is currently developing an end-to-end prototype 
of the CTA SST to be installed, commissioned  and operated under field conditions 
in Italy during 2014. 
The telescope, named ASTRI SST-2M, is characterized
by two innovative technological solutions, for the first time
adopted together in the design of Cherenkov telescopes: the
optical system is arranged in a dual-mirror Schwarschild-Couder (SC)~\cite{bib:SC} configuration \cite{bib:CANESTRARI}
and the camera at the focal plane is composed by a matrix of multipixel silicon photo-multipliers (SiPMs) 
\cite{bib:CATALANO}.

Current Cherenkov telescopes adopt parabolic or Davies-Cotton optical configurations with $f/D \gtrsim 1$. 
The plate scale of such telescopes is of the order of 25~cm/deg,  which is
well suited to cameras composed by about one inch photo-multiplier tubes 
(adopted by all present Cherenkov telescope arrays)  
corresponding to the optimal pixel size of $0.1^{\circ}-0.2^{\circ}$.  
Using a double mirror optics in SC configuration 
the plate scale can be reduced to 30-40~mm/deg, 
allowing the use of much smaller devices like the innovative silicon photo-multipliers 
whose typical size is few millimeters.
Moreover an optical point spread function smaller than the pixel size over a large field of view 
$\sim 10^{\circ}$ can be obtained with a much more compact SC telescope than a traditional one.
The silicon photo-multiplier 
technology is developing very quickly. 
The very high dark count rate and optical cross talk affecting old SiPMs have been rapidly decreasing in the last years 
as well as the unit cost, while the photon detection efficiency has been steadily increasing. 
Cherenkov telescopes equipped with such photon detectors, 
which are much less sensitive to high light flux than traditional photo-multipliers, 
can observe  even with full Moon increasing their observing time with respect to the current telescopes 
and providing a better coverage of variable emissions that more and more frequently appear in the VHE range. 
Also the measurement of the $e^{+}/e^{-}$ ratio in cosmic rays \cite{bib:COLIN}, 
that Cherenkov telescopes can obtain 
observing the Moon shadow could be improved using such devices.
The SiPMs have been already successfully used for the camera of the FACT Cherenkov telescope \cite{bib:FACT}, 
operating at La Palma since October 2011 but they have never been used so far for the camera of a SC telescope.  
FACT is a traditional Davies-Cotton telescope and therefore needs light concentrators to focus the photons
reaching its focal plane onto the SiPMs. 
The use of a secondary mirror, as planned for the ASTRI-SST-2M telescope, instead of light cones 
could further improve the performance of the telescope. 
The CTA consortium foresees the construction of two arrays, 
one in the Northern hemisphere  and a second one in the Southern hemisphere. 
The sites of the two arrays have not been decided yet, so the ASTRI collaboration decided to test its SST prototype 
at the INAF ”M.G. Fracastoro” observing station \cite{bib:MACCARONE} located in Serra La Nave, 1735~m a.s.l. 
on the Etna Mountain near Catania, Italy.
The construction, commissioning and the first observational campaigns will 
allow the test of all the telescope components (mechanics, mirrors, camera, front-end electronics) 
in real observing conditions.  
Obviously the final test will be the detection of real VHE sources, like the Crab Nebula. 
The expected sensitivity has therefore to be determined
to estimate the observing time needed to detect such sources.
In this paper an accurate Monte Carlo (MC) simulation has been developed and the results on the  
ASTRI-SST-2M prototype energy threshold, sensitivity and angular resolution are presented and discussed.

\section{The ASTRI-SST-2M telescope prototype}

The telescope design, 
fully compliant with the CTA requirements for the SST array, 
is very compact,   with a 4.3~m diameter primary mirror and a 1.8~m diameter 
secondary mirror at 3.1~m distance \cite{bib:CANESTRARI}. 
The telescope implements the so-called Schwarzschild-Couder configuration, 
an aplanatic, wide field of view, double reflection optical layout. 
The primary mirror is tessellated into 18 facets
distributed in three concentric coronae,
while the secondary mirror is a monolithic element.
The SC optical design has an f-number $f/0.5$ resulting in a plate scale 
of 37.5~mm/deg and an equivalent focal length of 2.150~m.
The focal plane is curved with a curvature radius of 1~m.  
The telescope mount exploits the classical alt-azimuth configuration. 
The light is focused on a compact camera \cite{bib:CATALANO} using multipixels SiPMs (Hamamatsu S11828-3344M) ) 
as photon detection devices with 1984 logical pixels, $6.2~mm \times 6.2~mm$, 
(obtained coupling 4 physical pixels, $3~mm \times 3~mm$),
corresponding to an 
angular size of $0.17^{\circ}$, very close to the optimal value. 
The SiPM signals are processed by  a front-end electronics using 
the EASIROC ASICs \cite{bib:EASIROC}. 
%
\section{Monte Carlo simulations}

\begin{figure}[htb]
  \centering
  \includegraphics[width=0.48\textwidth]{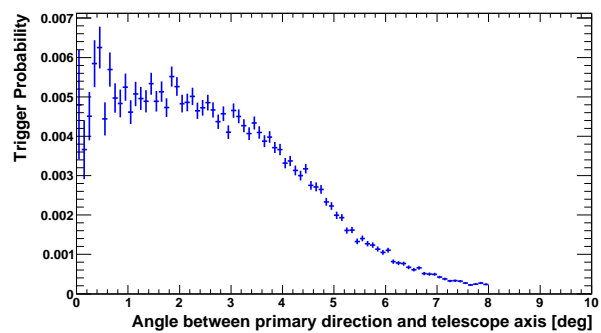}
  \caption{Trigger probability for proton showers as a function of the angle between the primary direction and the telescope axis.}
  \label{TRIGGERPROB}
\end{figure}

The standard CTA package for the simulation of the shower development in the atmosphere 
(CORSIKA version 6.99 \cite{bib:CORSIKA}) has been used 
with the IACT plugin package 
to randomize the shower 
core many times and save only the Cherenkov photons hitting the observational level in a region close 
to the telescope position. In this way the simulation of the large number of showers needed to estimate the 
telescope sensitivity can be increased while the needed disk-space is reduced.  
Showers produced by gamma primaries from a point-like source at $20^{\circ}$ zenith angle 
and diffused protons at the same zenith angle 
within a cone with $8^{\circ}$ radius have been simulated.     
Such a big diffusion angle for protons is necessary to correctly take into account the background contribution 
of events which  
are detected by the telescope even if they are quite outside of the telescope full FOV of $9.6^{\circ}$.
Figure \ref{TRIGGERPROB} shows the trigger probability for proton events as a function of the angle between 
the primary direction and the telescope axis. It is worthwhile to notice that there is still a 10\% relative 
trigger probability for primaries at angles larger than $5^{\circ}$.
The spectral index used for both primaries is $-2.0$ 
(rescaled to $-2.62$ for gammas and to $-2.7$ for protons in the analysis step) 
to equally distribute the CPU time 
over the entire energy range from 300~GeV to 100~TeV.  
No showers induced by heavier nuclei nor by electrons have been generated. 
The telescope response has been simulated with the sim\_telarray package \cite{bib:SIMTELARRAY} for what concerns 
the photon propagation from the primary mirror
to the secondary one till the pupil of the focal plane camera,
the photo-detection by SiPMs and the trigger logic. 
The presence of 4 adjacent pixels with a signal above 3.2 photoelectrons has been imposed as trigger condition. 
The simulation of the electronic chain  
has been performed with a custom code to properly take into account 
the characteristics of the EASIROC chip. 
The main characteristics of the simulated showers are summarized in table \ref{MCsamples}. 
\begin{table}[h]
\begin{center}
\begin{tabular}{|l|c|c|}
\hline                         & Gamma    &  Proton    \\ \hline
Events                         &   5.9E5  &   5.6E6    \\ \hline
Maximum impact point [m]       &   1500   &  2000      \\ \hline
Core  positions                &     10   &    20      \\ \hline
\end{tabular}
\caption{Monte Carlo samples.}
\label{MCsamples}
\end{center}
\end{table}
The simulation of the large amount of needed atmospheric showers has been realized exploiting the GRID technology
\cite{bib:GRID}. 

\section{Analysis of simulated data }

\begin{figure}[htb]
  \centering
  \includegraphics[width=0.48\textwidth]{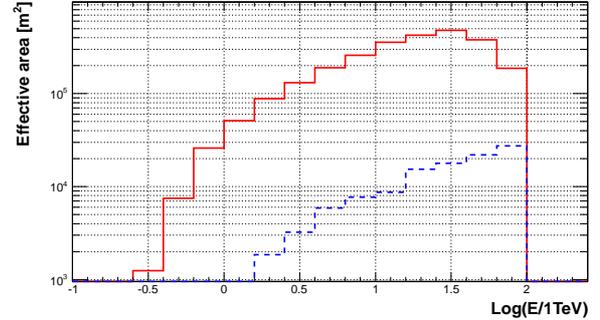}
  \caption{Effective area for gammas (red solid line) and effective area for protons times 100 (blue dashed line) 
after the gamma/hadron discrimination cuts.}
  \label{EFFECTIVEAREA}
\end{figure}

\begin{figure}[htb]
  \centering
  \includegraphics[width=0.48\textwidth]{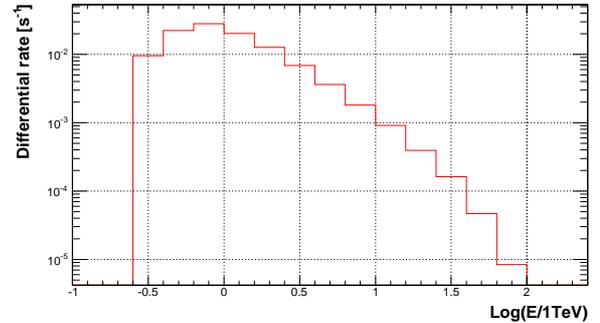}
  \caption{Expected differential rate from the Crab Nebula.}
  \label{CRABRATE}
\end{figure}

\begin{figure}[t]
  \centering
  \includegraphics[width=0.48\textwidth]{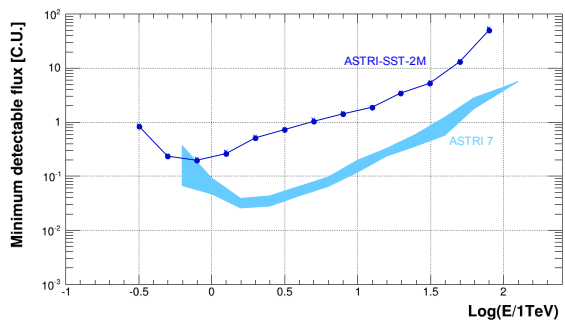}
  \caption{The expected differential sensitivity of the ASTRI-SST-2M prototype in Crab Units (C.U.) (blue points).
    The expected sensitivity of the ASTRI mini-array with 7 telescopes is shown for comparison (blue shaded area) \cite{bib:FEDERICO}.}
  \label{SENSITIVITY}
\end{figure}

The simulated data have been analyzed with the Eventdisplay package \cite{bib:EVNDISP}. 
Shower images have been cleaned with the robust two fixed levels cleaning algorithm:
only pixels with a signal above a certain threshold are considered to look for 
adjacent pixels above a lower threshold. We used 13 photoelectrons for the higher 
threshold and 7 photoelectrons for the lower one to get rid of the Night Sky Background contribution
(the simulated NSB rate is 24~Mhz per pixel). 
The resulting images have been selected requiring at least 4 pixels and an overall signal, the so-called size, 
greater than 50 photoelectrons. 
Images surviving to these preliminary cuts have been parameterized using the standard second moment Hillas analysis. 
For the gamma/hadron separation we adopted the so-called supercuts, firstly introduced in the Whipple data analysis, 
as described in \cite{bib:SUPERCUTS}. 
This method is surely not the best performing but it is very robust and well suited 
for a first estimate of the telescope performance. 
The adopted cuts on the Hillas parameters are summarized in table \ref{CUTS}. 
\begin{table}[h]
  \begin{center}
    \begin{tabular}{|l|c|c|}
      \hline Parameter & Min value  & Max value   \\ \hline
      Length    & $0.10^{\circ}$  & $1.00^{\circ}$    \\ \hline
      Width     & $0.01^{\circ}$  & $0.20^{\circ}$ \\ \hline
      Dist      & $0.50^{\circ}$  & $3.50^{\circ}$  \\ \hline
      Asymmetry & 0.0           &               \\ \hline
      Alpha     &    &  $10^{\circ}$   \\ \hline
    \end{tabular}
    \caption{Cuts used to reduce the hadronic background.}
    \label{CUTS}
  \end{center}
\end{table}
About $52\%$ of gamma events and $2\%$ of proton events 
survived to all these cuts with 
an enhancement of the quality factor, $S/\sqrt(B)$ ratio, by 3.7. 
With these data we estimated the effective areas for gammas and protons which are shown in figure \ref{EFFECTIVEAREA}.  
It is worth to notice that the effective area for gamma events drops slightly at energies greater than 30~TeV due 
to the image length of such energetic showers which is so large to not fulfill the cuts on this parameter. 
This effect could be reduced with more refined selection cuts, like dynamical cuts, i.e. cuts depending 
on the image size \cite{bib:DYNAMICCUTS}, 
or with much more advanced analysis techniques like the Random Forest 
successfully used by the MAGIC collaboration \cite{bib:RANDOMFOREST}.   
The expected rate of events from the Crab Nebula can be calculated convolving the it's spectrum 
\begin{eqnarray*}  
  \Phi_{Crab} = 2.83 \cdot 10^{-7} m^{-2} s^{-1} TeV^{-1}  (\frac{E}{1 TeV})^{-2.62}   
\end{eqnarray*}
\cite{bib:CRABSPECTRUM} with 
the effective area (see figure \ref{CRABRATE}). Integrating the obtained differential rate 
over the entire energy range we can calculate 
that about 75 events per hour 
will be detected by the ASTRI-SST-2M telescope prototype.     
The energy threshold, defined as the peak of the Crab Nebula differential rate is about 800~GeV.
The differential sensitivity of the telescope has been calculated assuming 50 hours observation time on-axis and 
requiring a significance of at least 5 sigmas in each energy bin (calculated accordingly to the 
Li\&Ma formula 17 \cite{bib:LIMA}), at least 10 gamma events and a gamma-ray rate after all cuts greater than five percent 
of the background rate. The expected number of background events in each bin has been calculated 
assuming a power law spectrum for proton events \cite{bib:PROTONFLUX}: 
\begin{eqnarray*}
 \Phi_{proton} = 0.096 \cdot  10^{-7} m^{-2} s^{-1} sr^{-1} TeV^{-1}  (\frac{E}{1 TeV})^{-2.7}   
\end{eqnarray*}
The differential sensitivity of the ASTRI-SST-2M prototype is shown in figure \ref{SENSITIVITY} compared with 
the differential sensitivity expected for the ASTRI mini-array with 7 telescopes.

The location of the gamma source on the focal plane has been determined using the Disp method, 
as described in \cite{bib:DISP}. The position of the source must lie on the major axis of the image
in the direction indicated by the asymmetry of the image itself. The distance of the source from 
the image centroid, the so-called disp,  and the elongation of the image depend upon the impact parameter 
of the shower at the observational level. As the impact parameter grows the disp grows 
as well as the elongation of the image.
The latter can be expressed as the ratio of image angular width and length. 
At the first order a linear relationship between the disp parameter and the image elongation can be assumed: 
\begin{eqnarray*}
 disp =  \xi \left ( 1 -\frac{width}{length} \right )   
\end{eqnarray*}
where $\xi$ is a scaling parameter to be determined with MC simulations. 
A combination of these features provides a unique arrival direction for each gamma-ray event. 
The optimal value for the scaling factor $\xi = 1.23^{\circ}$ has been determined choosing 
the value which minimizes the sigma of a bi-dimensional distribution of the source position points 
obtained with a MC sample of gamma showers. 
This value has been used to analyze an independent sample to measure the angular resolution  defined as the radius 
containing 68\% of the events. 
The resulting angular resolution of the ASTRI-SST-2M prototype over the entire energy range is $ 0.49^{\circ}$.

\section{Discussion and Conclusions}

The construction of a prototype telescope for the SST array is certainly of primary importance for testing 
all its components,  mechanics,  mirrors,  camera  and electronics.
Its operation will show the response and reliability of the various subsystems in real working conditions
and will allow the measurement of its optical and pointing capabilities. 
Also the calibration procedure, related to the pixel response and the optics alignment could be checked. 
The study described in this paper shows moreover that the ASTRI-SST-2M telescope will be able to detect 
gamma-rays with an energy threshold $E_{Th} \sim 800~GeV$ 
and a mean angular resolution of $0.49^{\circ}$. 
These values are not far from the same figures of merits of the 
 Whipple telescope, 300~GeV and $0.3^{\circ} @ 300~GeV$ respectively. 
It is worthwile to underline that the present results have been obtained with a very robust analysis 
based on super-cuts for the gamma/hadron separation and on the Disp method for the determination of the source position. 
There are other well-known methods with better performance which make us confident that 
there is still much room for improvements.  
The energy resolution of the ASTRI-SST-2M telescope has not been estimated yet. It will be for sure worse than 
that of present telescope arrays because with one telescope it is hard to disentangle the effect of the 
primary energy and impact paramenter on the image size. Moreover, the ASTRI-SST-2M telescope thanks to the large 
field of view  will detect events with very large core distances $\sim 300~m$, whose image size depends strongly on
the core distance. Even a small uncertainty on the core distance will result in a large uncertainty on the energy of such events.   
The limited sensitivity ($\sim 0.2$~C.U. at 800~GeV) will not allow any competition with respect to the existing arrays 
and only the detection of the Crab Nebula and of the brightest AGNs (MRK 421 and 501) is foreseen.  
The Crab Nebula can be detected at 5~sigma level in about 2 hours, while Mrk~421 and Mrk~501 in low state (0.30 C.U. and 0.25 C.U.) 
in 22 and 32 hours respectively. 
The second step of the ASTRI project will be the installation on the selected CTA Southern site 
of a mini-array of 5-7 SST telescopes \cite{bib:FEDERICO}. 
From this configuration the performance in terms of threshold energy, 
angular and energy resolution and point source sensitivity could be easily extrapolated to the whole SST array, 
making us confident on the CTA performance. 

\vspace*{0.5cm}

\footnotesize{{\bf Acknowledgment:}{
We gratefully acknowledge support from the agencies and organizations   
listed in this page: http://www.cta-observatory.org/?q=node/22.
This work was partially supported by the ASTRI Flagship Project financed by the Italian Ministry of Education,
University, and Research (MIUR) and led by the Italian National
Institute of Astrophysics (INAF).
}}
\end{document}